# Predicting nonlinear reshaping of periodic signals in optical fibre with a neural network

**Sonia Boscolo** [1], **John M. Dudley** [2] **and Christophe Finot** [3,*]

[1] *Aston Institute of Photonic Technologies, School of Engineering and Applied Science, Aston University, Birmingham B4 7ET, United Kingdom*

[2] *Institut FEMTO-ST, UMR 6174 CNRS-Université de Franche-Comté, Besançon, 25000, France*

[3] *Laboratoire Interdisciplinaire Carnot de Bourgogne, UMR 6303 CNRS-Université de Bourgogne, 9 avenue Alain Savary, 21078 Dijon Cedex, France*

christophe.finot@u-bourgogne.fr

**Abstract:** We deploy a supervised machine-learning model based on a neural network to predict the temporal and spectral reshaping of a simple sinusoidal modulation into a pulse train having a comb structure in the frequency domain, which occurs upon nonlinear propagation in an optical fibre. Both normal and anomalous second-order dispersion regimes of the fibre are studied, and the speed of the neural network is leveraged to probe the space of input parameters for the generation of custom combs or the occurrence of significant temporal or spectral focusing.

# I. Introduction

The accumulation of nonlinear effects in an optical fibre is often seen as a source of significant impairment for the propagating light signals, but the same effects, when properly managed, can provide a remarkable tool to tailor the temporal and spectral content of the signals. Indeed, depending on the regime of dispersion of the fibre and the frequency chirp, an initial pulse can be significantly expanded or compressed in the time or frequency domain, or it can be reshaped into advanced temporal waveforms such as parabolic, rectangular and triangular shapes [1]. Yet, due to the typically wide range of degrees of freedom involved, predicting the behaviour of nonlinear pulse shaping by numerical integration of the nonlinear Schrödinger equation (NLSE) or its extensions may be computationally demanding, especially when dealing with inverse-mapping problems. Recently, we have successfully introduced the use of the machine-learning (ML) method of artificial neural networks (NNs) as an efficient tool for complementing or substituting the NLSE in the modelling of nonlinear pulse shaping [2-5] or for predicting the generation of optical supercontinua [6, 7].

Fibre nonlinearity does not only affect the propagation of ultrashort pulses. A continuous wave modulated at a frequency $f_m$ will also experience energy exchange between the spectral lines that make up its spectrum, along with a change in the relative phase between the frequency components. New equally spaced frequency components will emerge giving rise to a frequency comb. Concomitantly, significant reshaping will take place in the time domain, potentially leading to very high repetition rate trains of ultra-short pulses [8]. The essential dynamical features of this four-wave mixing (FWM) process can be understood from a truncated model involving three or four frequency components [9-12]. Very recently, we have showed the benefits of ML in addressing this problem by demonstrating that a NN can be readily trained from experimental data to accurately model the wave mixing process [13], and that data-driven discovery based on sparse regression can be used to extract the governing differential equation model [14]. Nevertheless, the validity of the truncated model is limited by the emergence of higher-order spectral sidebands that typically occurs in practice due to further wave mixings. Semi-analytical models involving a higher number of frequency components have been explored [15-17], but their complexity and underlying assumptions may restrict their practical implementation. In this paper, we implement an artificial NN to predict the spectro-temporal evolution of a periodic waveform signal upon its nonlinear propagation in an optical fibre. We demonstrate the ability of the trained NN to identify the initial conditions that are required to generate flat frequency combs or optical spectra where one frequency component is cancelled in the fibre. The longitudinal evolution of the wave intensity profiles in the time and frequency domains are accurately predicted by the NN for both the anomalous and normal dispersion regimes of the fibre, which therefore enables efficient and accurate replication of the processes of ultrashort pulse formation, spectral compression and undular bores that arise from NLSE dynamics.

## II. Problem under study and modelling tools

### A/ Physical model

The general problem studied in this paper is the nonlinear propagation of two types of periodic waveforms in an optical fibre, as illustrated in Fig. 1. By limiting our analysis to symmetrical initial conditions, we can mathematically describe the optical spectral field $\tilde{\psi}(0,\omega)$ with two parameters only, namely the ratio $A^2$ of the intensity of the central frequency component to the intensity of the lateral sidebands, and the spectral phase $\varphi$ of the lateral sidebands relative to the central component. Here, $\tilde{\psi}(z,\omega)$ is the slowly varying envelope of the field in the frequency domain. Such initial wave conditions have already been studied in the context of linear shaping and shown to give rise to the synthesis of various temporal waveforms according to the values of $A^2$ and $\varphi$ [18]. In one case, a continuous wave is modulated at the angular frequency $\omega_m = 2\pi f_m$, yielding an optical spectrum made of a central component and two sidebands at the frequencies $\pm f_m$,

$$\tilde{\psi}(0,\omega) \propto \delta(\omega) + A \exp(i\varphi) \left[\delta(\omega-\omega_m) + \delta(\omega+\omega_m)\right], \tag{1}$$

The corresponding temporal intensity profile is given by

$$I(0,t) \propto 1 + 2A^2 + 4A\cos(\varphi)\cos(\omega_m t) + 2A^2 \cos(2\omega_m t), \tag{2}$$

The other case refers to four spectral lines with no continuous background, where the spectral field amplitude can be expressed as

$$\tilde{\psi}(0,\omega) \propto \delta\left(\omega-\frac{\omega_m}{2}\right)+\delta\left(\omega+\frac{\omega_m}{2}\right)+A\exp(i\varphi)\left[\delta\left(\omega-\frac{3\omega_m}{2}\right)+\delta\left(\omega+\frac{3\omega_m}{2}\right)\right], \quad (3)$$

leading to the following temporal intensity profile:

$$I(0,t) \propto \begin{Bmatrix} 1+A^2+\left[1+2A\cos(\varphi)\right]\cos(\omega_m t)+ \\ 2A\cos(\varphi)\cos(2\omega_m t)+A^2\cos(3\omega_m t) \end{Bmatrix}, \quad (4)$$

We neglect initial quantum noise in our model, so that we can neglect the growth of spontaneous modulation instability upon propagation in the anomalous regime of dispersion.

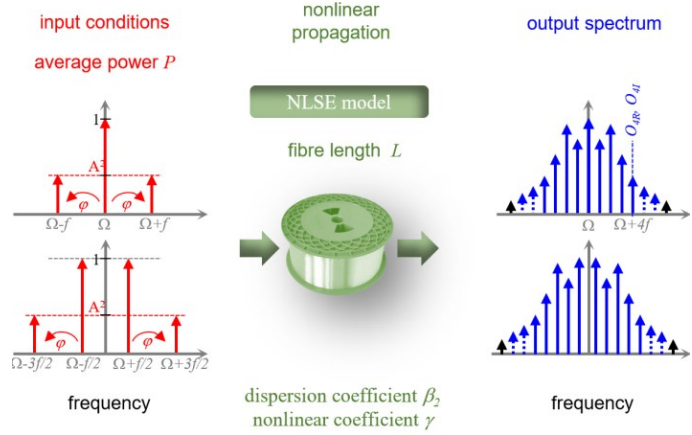

Figure 1: Schematic of the nonlinear shaping problem under investigation.

The evolution of the slowly-varying scalar field envelope $\psi(z,t)$ upon propagation in a lossless, single-mode fibre is governed by the nonlinear Schrodinger equation (NLSE) [19]:

$$i\frac{\partial\psi}{\partial z}-\frac{\beta_2}{2}\frac{\partial^2\psi}{\partial t^2}+\gamma|\psi|^2\psi=0, \quad (5)$$

where $z$ is the propagation distance, $t$ is the time in a reference frame travelling at the group velocity, and $\beta_2$ and $\gamma$ are the second order (group-velocity) dispersion and Kerr nonlinearity parameters of the fibre, respectively. It is useful to normalise Eq. (5) by introducing the dimensionless variables: $u = \psi/\sqrt{P_0}$, $\xi = z/L_D$, $\tau = t/T_0$, and write it in the form

$$i\frac{\partial u}{\partial \xi} - \frac{\text{sgn}(\beta_2)}{2}\frac{\partial^2 u}{\partial \tau^2} + N^2 |u|^2 u = 0 \qquad (6)$$

where $P_0$ is the average power of the input modulated wave, $L_D = 1/(|\beta_2| f_m^2)$ and $L_{NL} = 1/(\gamma P_0)$ are the respective dispersion length and nonlinear length associated with the initial waveform, and the power parameter (or soliton number) $N$ is defined by $N^2 = L_D/L_{NL}$. This way, the nonlinear shaping problem, which depends on the seven physical parameters ($f_m$, $A$, $\varphi$, $P_0$, $\beta_2$, $\gamma$, $L$), where $L$ is the fibre length, is mapped onto a problem in the four-dimensional space of ($A$, $\varphi$, $\xi$, $N$). This dimensionality reduction relaxes the complexity of the problem, and for a specific selected set of normalised parameters there are many groups of physical parameters suitable the defining equations of $A$, $\varphi$, $\xi$ and $N$.

The variation of the initial spectral phase offset $\varphi$ can be restricted to an interval of length $\pi$ rad, hence, we have used values between $-\pi/2$ and $\pi/2$ rad. The time-domain results for other values of $\varphi$ can then be easily retrieved using half-period temporal shifts. For the NLSE to remain fully valid without significant influence of higher-order terms, we have adapted the ranges of variation of the other three input parameters to the fibre dispersion regime and initial wave condition considered, as summarised in Table 1. Negative values of the amplitude ratio $A$ relate to cases when the central frequency component of the initial wave is less intense than the surrounding sidebands. Typically, for a fibre with anomalous dispersion, we have explored $N$ values between 2 and 10 and propagation lengths up to the minimum value between 1.5 $L_{NL}$ and 0.2 $L_D$. Within such a parameter span, we do not expect to observe any recurrence wave patterns, where this phenomenon is not the target of the present discussion. Propagation in the normal dispersion regime is known to require higher powers and longer distances. Therefore, in this case $N$ is varied between 2 and 20 and the maximum propagation length is set to the minimum value between 4 $L_{NL}$ and 0.5 $L_D$. We note that the input parameter span explored is large, well beyond that typically used in related publications: the relative intensity of the sidebands $A^2$ may vary by a factor between 100 and 2000, while the average signal power may change by a factor of 25 or 100.

| Type of initial waveform | Dispersion regime | $A$ (dB) | $\xi_{max}\ L_D$ | $N$ |
|---|---|---|---|---|
| 3 spectral lines | anomalous | [-3 ; 15] | min(.2 $L_D$ ; 1.5 $L_{NL}$) | [2 ;10] |
| 3 spectral lines | normal | [-3 ; 15] | min(.5 $L_D$ ; 4 $L_{NL}$) | [2 ;20] |
| 4 spectral lines | anomalous | [7 ; 17] | min(.2 $L_D$ ; 1.5 $L_{NL}$) | [2 ;10] |
| 4 spectral lines | normal | [2 ; 20] | min(.5 $L_D$ ; 4 $L_{NL}$) | [2 ;20] |

Table 1: Ranges of input parameters investigated relative to the initial waveform type and fibre dispersion

Contrary to most of the ML-based approaches recently applied to NLSE propagation problems, here we do not sample the simulation output wave profiles along both the temporal and spectral dimensions, which would require a significant number of sampling points (above 100 points were used in [4]), but we focus instead on the signal spectrum. Given the periodicity of the signal being studied, the spectrum has a comb structure and, thus, recording a limited number of output complex spectral amplitude values is enough to accurately reproduce all the features of the waveform. Typically, and exploiting the symmetry of the problem, we have recorded the complex amplitude of the five most central spectral lines with zero or positive frequency. Rather than memorising the complex amplitude in terms of modulus and phase (which may incur a $2\pi$ ambiguity), we have stored the real and imaginary parts separately. Therefore, the results of the wave propagation can be represented by only ten data points. This approach, which is suited to the periodic waves being studied, represents an important simplification of the problem compared to existing strategies, while it still enables access to all the amplitude and phase features of the signal in both the time and frequency domains.

## B/ Artificial neural network and optimum solution search

We have deployed a classic multilayer perceptron in our study, as shown in Fig. 2(a). It is worth noting that recently, many works have explored the application of more advanced ML methods to NLSE-related problems, such as recurrent [5, 6, 20], convolutional [21-24], or physics-informed [25-28] NNs, achieving remarkable performance. In this work, our goal is not to showcase the most powerful ML method for solving the NLSE propagation problem being considered but rather to emphasise that an easily accessible and model-free method [29] can already fit our purpose well. As described in the previous section, the input parameters for the NN are ($A$, $\varphi$, $\xi$, $N$) while the output parameters are the real and imaginary parts of five spectral components of the signal after propagation in the fibre. A rather large data set is required for training the NN. Therefore, we have solved the NLSE - using a standard split-step Fourier propagation algorithm - for $2\times10^5$ randomly chosen combinations of initial conditions covering the ranges shown in Table 1. 75% of the resulting data are used to train the NN while the rest is kept for validation. The NN structure consisting of three hidden layers with thirty-six, thirty and twenty-six neurons, respectively, was chosen in an empirical manner. We also tested a cascaded forward NN but the slight gain in accuracy did not justify the significant additional cost in training time. The Bayesian regularisation back propagation algorithm was used for training, and the NN was programmed in Matlab using the neural network toolbox.

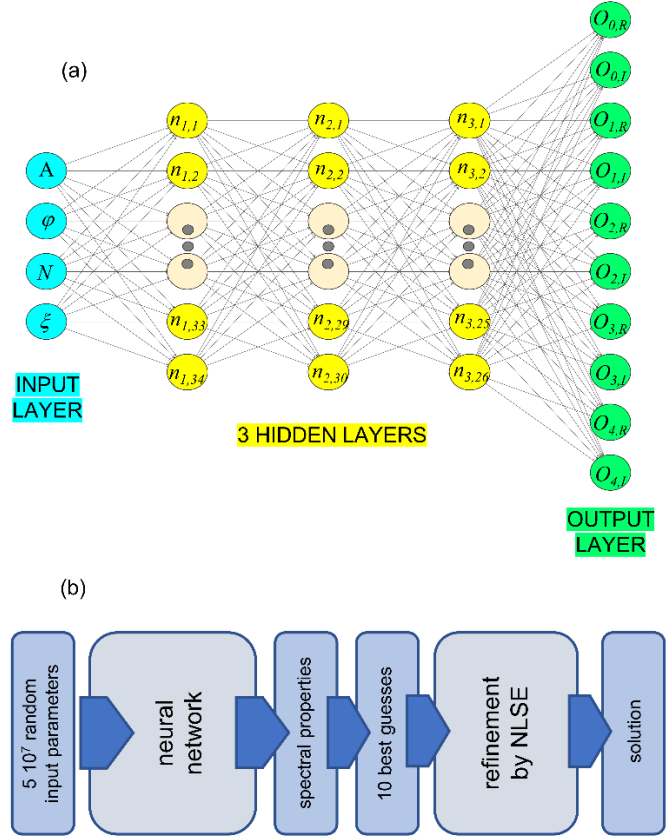

Figure 2: (a) NN architecture deployed. (b) Principle of searching for the optimum solution.

Our aim is to get the trained NN able to accurately predict the temporal and spectral features of the output waveform when tested on new data, with a computation time several orders of magnitude below the time required by the usual numerical integration of the NLSE. With such a speed indeed, we will be able to search the full four-dimensional space of input parameters for a solution that may meet a given target, such as a frequency comb fulfilling a predefined spectral pattern or the generation of a pulse train with the shortest pulse duration. We would like to emphasise that contrary to [3, 4], where we demonstrated that a properly trained NN can identify with very high accuracy the nonlinear propagation and initial pulse properties from the pulses observed at the fibre output, here we do not know a priori whether the targeted waveform can be physically produced. We have employed a very simple search process (Fig. 2(b)): we test the trained NN on $5 \times 10^7$ new simulations with randomly chosen initial conditions which are sampled from the input parameter space with a rather high density. Remarkably, the NN can predict the output spectral properties corresponding to this large data set in less than a minute (40 seconds on a personal computer with a 10-core processor, 3.70-GHz processor frequency, and 64-GB memory). We then evaluate the ten best guesses from these spectral properties (or temporal properties recovered by a simple Fourier transform) and, as the NN predictions may be affected by very small inaccuracies, we lastly refine the solution by running a few hundred additional NLSE numerical simulations in the very close vicinity of the best guesses. The solution that is the closest to the targeted output waveform is then retained. We would like to emphasise that our goal is not to benchmark our approach against well-established methods for optimisation and search problems such as, e.g., genetic algorithms [30, 31] or gradient descent methods [32].

## III. Generation of custom frequency combs

In this section, we study the problem of generating frequency combs with pre-defined profiles through multiple FWM interactions in the fibre.

### A/ Flat combs

We first target the generation of flat-top combs. Various works have studied the problem of generating a flat comb from a continuous wave and proposed solutions that, for example, take advantage of the combination of phase and intensity modulation [33]. Here, we investigate the nonlinear propagation in a fibre with anomalous dispersion of a three frequency-component initial waveform (Eq. (1)). Our target is a symmetric comb made of nine spectral lines of equal amplitude. Using the NN architecture described in the previous section, we are able to test millions of input parameter combinations in our four-dimensional space and evaluate for each combination the ratio $C$ of the spectral intensity of the most intense component to the intensity of the least intense component: $C = \max(I_0, I_1...I_4) / \min(I_0, I_1...I_4)$, where $I_i$ is the intensity of the $i^{th}$ component. The output frequency comb corresponding to the best combination of input parameters (i.e., the one yielding the $C$ value closest to 1) is shown in Fig. 3. For the initial amplitude ratio $A = -0.6455$, phase offset $\varphi = 0.41$ rad, normalised propagation length $\xi = 0.0126$ and power parameter $N = 9.45$, the comb achieved by numerical integration of the NLSE (black diamonds) features intensity fluctuations between its components below 0.2 dB. The predictions from the NN (red circles) are in very good agreement with the NLSE results, although minor discrepancies can be observed, thereby justifying the final stage of refinement by NLSE simulations in our search procedure. It is worth mentioning that to avoid the occurrence of artifacts near the boundaries of the training dataset, we have used a margin of 5% with respect to the boundary values.

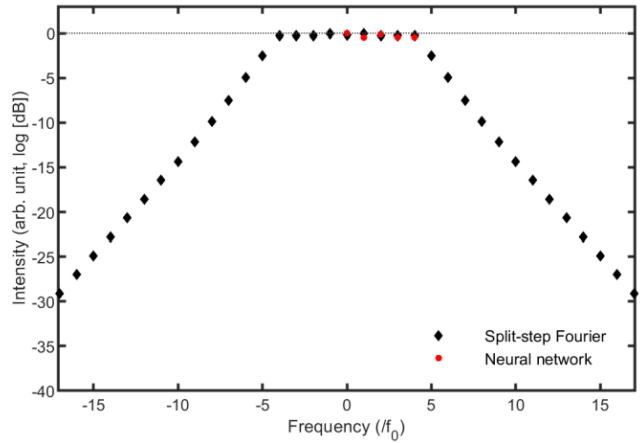

Figure 3: Generation of a comb consisting of nine spectral lines of equal amplitude through nonlinear propagation in an anomalously dispersive fibre. The predictions from the NN (red circles) are compared with the results of NLSE numerical simulations (black diamonds).

**B/ Flat combs with a component suppressed**

We have also explored the possibility of generating flat frequency combs with one of the components suppressed through propagation of a three-component initial waveform in an anomalously dispersive fibre. Previous works dealing with breathing structures [34] have demonstrated analytically and experimentally that it is possible to obtain full depletion of the frequency components of a continuous wave. However, the cascaded transfer of energy involved leads to a spectrum with a typical triangular shape when plotted on a logarithmic scale. Here, we try to achieve the cancellation of a frequency component while maintaining a high degree of flatness for the other components. Panels (a) and (b) of Fig. 4 shows examples of combs with the central component and the first lateral sidebands, respectively, suppressed, obtained for the respective sets of input parameters: ($A$ = 0.85 dB, $\varphi$ = 2.82 rad, $\xi$ = 0.0142, $N$ = 8.88) and ($A$ = 1.96 dB, $\varphi$ = 0.78 rad, $\xi$ = 0.0138, $N$ = 9.48). To achieve these targets, we have maximised the ratio of the average intensity $I_{av}$ of the three spectral components to retain to the intensity $I_c$ of the component to cancel. Here, a frequency component is deemed suppressed if the ratio of its intensity to the intensity of the most powerful component is below 5 $10^{-3}$ (i.e., –23 dB), in which case, its value in the cost function is replaced with 5 $10^{-3}$. The level of intensity fluctuations on the main part of the combs shown in Fig. 4 is below 0.4 dB. Once again, the predictions from the NN agree well with the results of NLSE simulations, hence substantiating the ability of the NN to act as an universal interpolator [35].

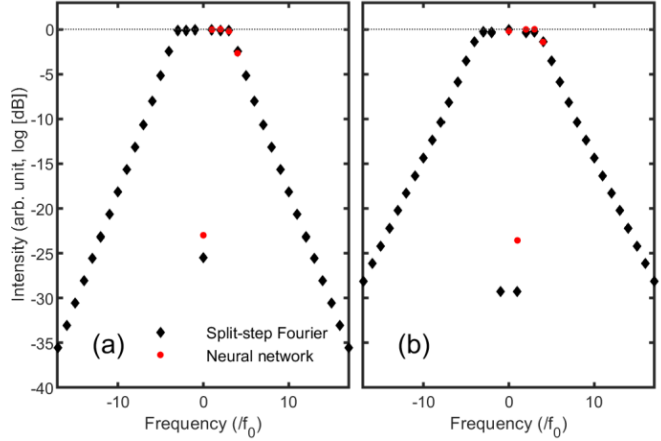

Figure 4: Generation of combs with (a) six spectral lines of equal amplitude and the central line suppressed, and (b) five spectral lines of equal amplitude and the first lateral sidebands suppressed, through nonlinear propagation in an anomalously dispersive fibre. The predictions from the NN (red circles) are compared with the results of NLSE simulations (black diamonds).

Our NN can also help reconstruct the evolution of the spectrum along the fibre length resulting into the formation of a targeted comb. Figure 5 relating to the comb shown in Fig. 4(b), shows that before the stage where the two lateral sidebands get cancelled, the central component experiences a stage of strong depletion. We can also observe the progressive broadening of the spectrum. However, given the finite number of output frequency components included in the NN, the network cannot obviously reproduce the outermost spectral lines. It is worth noting though that the evolution of the intensity of the frequency components predicted by the NN is in very close agreement with that obtained from integration of the NLSE. This is because the NN has been trained on a dataset based on the NLSE model, in other words, no assumption has been made, such as a truncated FWM model, that could affect the energy transfer among the reduced number of spectral sidebands considered in the NN.

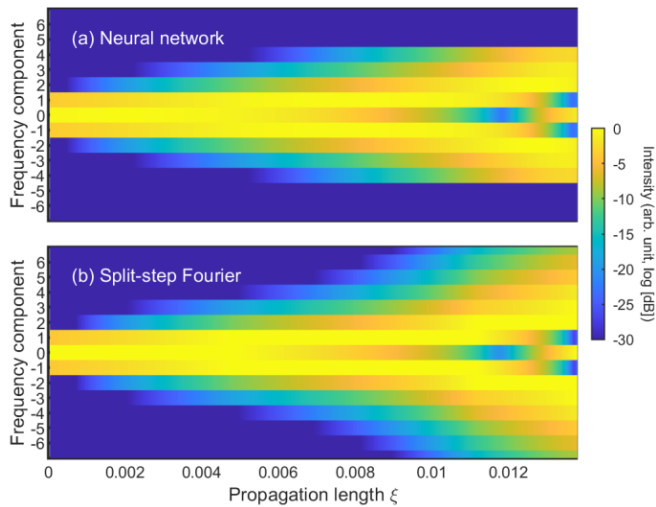

Figure 5: Longitudinal evolution of the optical spectrum into the frequency comb shown in Fig. 4(b). (a) Predictions from the NN. (b) Results of NLSE simulations.

## C/ Impact of initial conditions and dispersion regime

We have also tested the ability of the NN to search the input parameter space for the formation of flat combs starting from initial conditions (1) and (3) in both the anomalous and normal dispersion regimes of the fibre. The results obtained for each system's configuration are summarised in Fig. 6 where combs formed of seven or six spectral lines of equal amplitude are targeted. The associated input parameters are given in Table 2. It is noteworthy that, contrary to propagation in the anomalous dispersion regime, comb generation at normal dispersion requires in-phase initial conditions. In all cases, very flat combs are attained, featuring spectral wings on each side of the plateau which decay almost linearly when plotted on a logarithmic scale. While the intensity profiles in the frequency domain exhibit rather similar features for anomalous and normal dispersion, this is not the case in the time domain where we can note the very strong influence of the dispersion regime on the intensity profiles. Besides, the carrier-suppressed nature of the initial condition has a stronger impact on the output waveform in the normal dispersion regime.

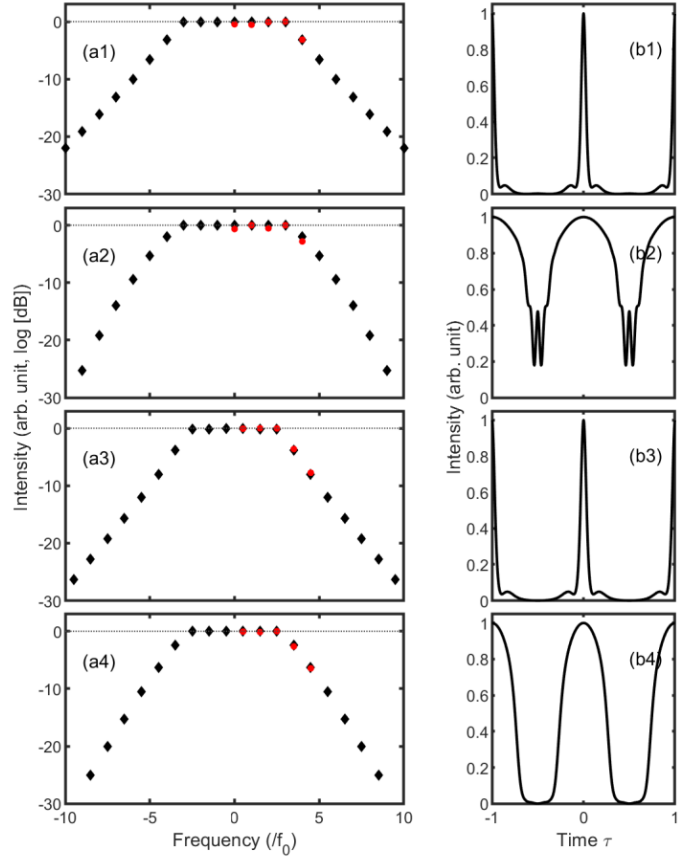

Figure 6: Generation of combs with six (panels 1, 2) and seven (panels 3, 4) spectral lines through nonlinear fibre propagation of three- and four-frequency component initial conditions, respectively. The results obtained in the anomalous and normal dispersion regimes of the fibre are shown in panels 1, 3, and 2, 4, respectively. (a) The optical spectra predicted by the NN (red circles) are compared with those obtained from NLSE simulations (black diamonds). (b) Corresponding temporal intensity profiles obtained from NLSE simulations.

| Type of initial waveform | Dispersion regime | $A$ (dB) | $\varphi$ (rad) | $\xi_{max}$ | $N$ |
|---|---|---|---|---|---|
| 3 spectral lines | anomalous | 2 | 0.216 | 0.0196 | 7.69 |
| 3 spectral lines | normal | 2.8 | 0.01 | 0.01 | 19 |
| 4 spectral lines | anomalous | 8.31 | 0.283 | 0.0283 | 7 |
| 4 spectral lines | normal | 3.12 | 0.005 | 0.0056 | 16.5 |

Table 2: Input parameters supporting the generation of the frequency combs shown in Fig. 6.

Inspired by [2], we have also built a visual representation of the input parameters that lead to the sought target. In Fig. 7, we therefore show the regions in the input parameter space that support values of the flatness parameter $C$ above 0.85 (corresponding to intensity fluctuations across the plateau of less than 0.7dB). The marker color in the figure is based on the value of the relative phase offset $\varphi$, and the results are plotted for the initial condition given by Eq. (1) and both regimes of dispersion. We can see that there are numerous combinations of input parameters that lead to the desired target. Strong differences between the two regimes of dispersion are also revealed. In the anomalous dispersion regime, two parameter branches are visible, relating to distinctly different $\varphi$ values. We also see that is possible to achieve a flat comb starting from both negative and positive values of the amplitude ratio parameter $A$. By contrast, in the normal dispersion regime the lateral spectral sidebands must be less intense than the central component, with a typical amplitude attenuation of around 3 dB. With respect to the relative phase offset, this must be kept very close to zero.

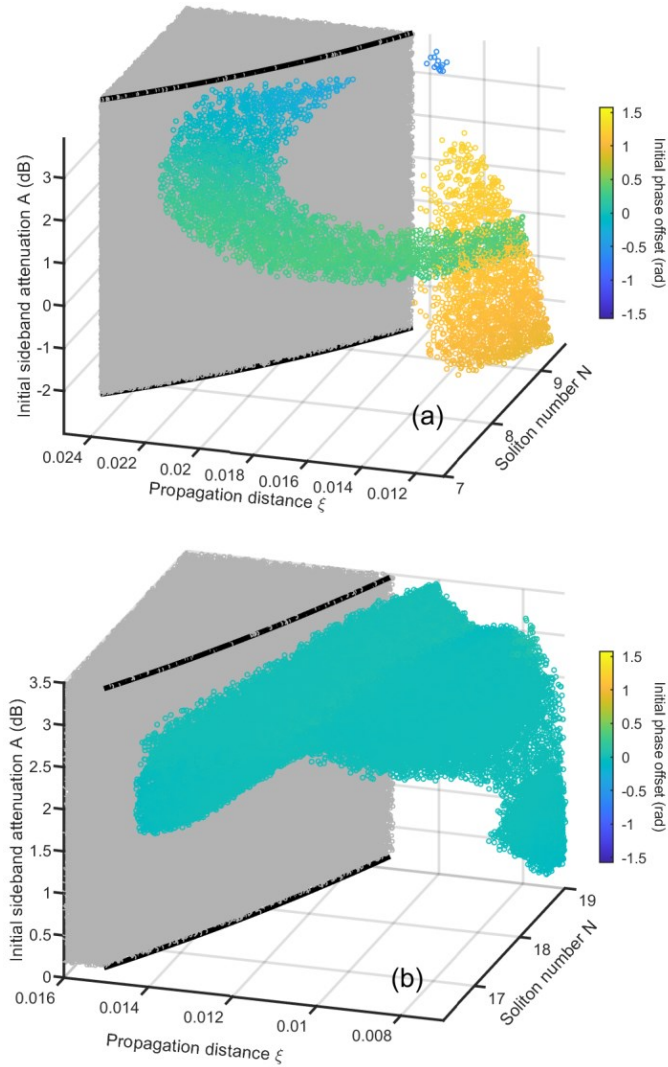

Figure 7: Regions in the space of input parameters ($A$, $\varphi$, $\xi$, $N$) that enable the formation of high-flatness combs starting from a three-frequency component initial condition. The results obtained for propagation in the anomalous and normal dispersion regimes of the fibre are shown in subplots (a) and (b), respectively. The data visualisation is restricted to the parameter volume where solutions are found, and the grey region represents volume that is unexplored.

## IV. Other examples of usage of the NN

In the previous section, we have discussed examples of tailoring the spectral intensity profile of the initial signal, where to achieve the target waveforms we could have deployed a NN only predicting the optical spectrum, hence, only having five output parameters. In this section, we address three additional problems that reveal the full strength of our NN, able also to deal with the temporal features of the signal.

### A/ Temporal compression process

The first practical problem that we address is the search of the input system's parameters that enable the generation of a train of pulses with the shortest duration, when we start from the initial wave condition given by Eq. (1). Impressive temporal compression is known to occur in the anomalous dispersion regime of the fibre [19], where modulation instability ensures efficient cascading of the FWM process. To achieve NN predictions in the time domain that are as accurate as possible, we have increased to seven the number of frequency components of the initial wave that are evaluated by the NN. From the fourteen output parameters given by the NN, we can reconstruct the symmetric spectrum made of thirteen spectral lines with known amplitude and phase, and then a simple inverse Fourier transform provides the periodic waveform in the time domain. We look for the most efficient temporal compression, which we define here as the generation of the pulse train with the highest peak to average power ratio. To limit our search to practically relevant, relatively low powers, we set the maximum value of the $N$ parameter to 6. Figure 8(a) shows the temporal intensity profiles of the initial and compressed waveforms for the optimum input parameters found by our search algorithm, namely, $A$ = 1.90 dB, $\varphi$ = 0.40 rad, $\xi$ = 0.0275, and $N$ = 5.79. We can see that starting from an initial wave with a duty cycle (defined as the ratio of the full width at half-maximum pulse duration to the pulse repetition period) of 1/3, the pulse duration is significantly decreased so that a duty cycle of 9% is achieved. Such a compression performance represents a significant improvement compared to the performance usually attained when starting from a simple sinusoidal waveform [8], which results in a typical duty cycle of 1/4. The logarithmic scale plot highlights that the intensity level of the surrounding pedestals is also extremely low (approximately –20 dB from the peak), ensuring that most of the energy is focused into the central part of the pulse. The predictions from the NN (red circles) are in excellent agreement with the results of direct NLSE simulations (black curves), thus confirming the confidence that we can have in the time-domain prediction capability of our NN. The comparison of the evolution of the temporal intensity profile of the initial wave along the fibre length obtained from the NN and NLSE simulations (Fig. 8(b)) further validates this conclusion.

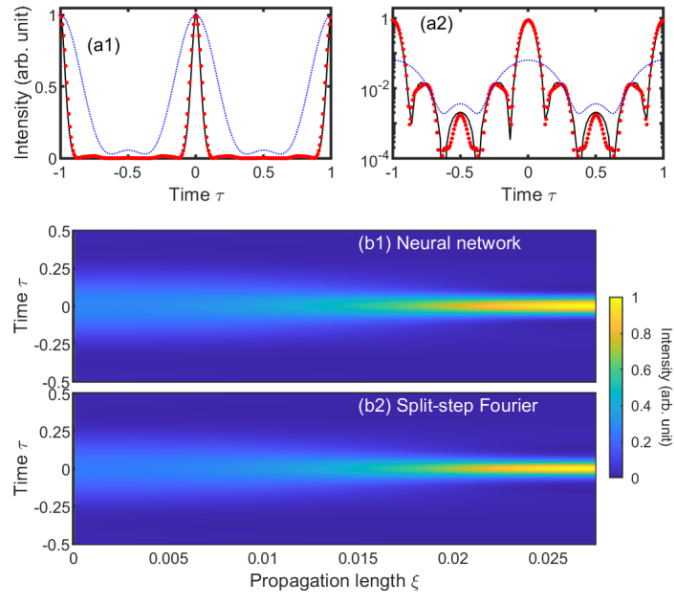

Figure 8: Pulse train with the highest peak power relative to the average power. (a) Temporal intensity profile plotted on linear (panel 1) and logarithmic (panel 2) scales. The predictions from the NN (red circles) are compared with the results of NLSE simulations (black curves). Also shown is the initial condition (dotted blue curves). The waveforms are normalised to the same peak power in panel 1 and to the same average power in panel 2. (b) Spatiotemporal evolution of the wave obtained from the NN (panel 1) and NLSE simulations (panel 2).

## B/ Spectral focusing process

The second problem that we study deals with nonlinear spectral focusing. Indeed, whereas self-phase modulation (SPM) is traditionally associated with broadening of the spectrum of a light signal and a concomitant energy transfer from the most powerful central frequency component to the neighbouring components, some system configurations may lead to the opposite phenomenon with a redistribution of the energy towards the central part of the spectrum [36, 37]. While SPM-induced spectral compression has been well studied for single pulses, only a few works have considered the case of continuous waves, where this spectral focusing phenomenon is also known as inverse FWM [38, 39]. Here, we consider an initial wave made of three spectral lines of equal amplitude. From a reduced FWM model in the anomalous dispersion regime [9, 13], one may anticipate that given the closed trajectory traced in the phase space, after a stage of spectral expansion, the wave will move back a slightly modulated wave. However, the resulting spectral compression is not very efficient. Therefore, here we consider the regime of normal dispersion where efficient spectral focusing is known to arise [40]. Figure 9(a) illustrates the space of input parameters that support a concentration of more than 80% of the initial energy into the central frequency component. We can see that the initial phase offset needs be around $\pm\pi/2$ and that a larger initial $N$ number enables a shorter propagation length. The optimum combination of parameters found is $\varphi = -1.32$ rad, $\xi = 0.025$, and $N = 11.5$. The related longitudinal evolution of the spectrum is shown in Fig. 9(b), which highlights the increasingly higher focusing of the energy into the central frequency component as well as the development of new and low-energy spectral sidebands.

The resulting compressed spectrum is shown in Fig. 9(c), highlighting that very efficient spectral focusing has taken place: while the three initial frequency components had the same intensity, the intensity of the lateral sidebands is now more than 13 dB below that of the central component. The two next neighbouring sidebands have similar intensity level. This spectral focusing is accompanied by significant reshaping in the time domain, where the initially sinusoidally modulated intensity profile is transformed into a train of rectangular-like pulses [40, 41]. The comparison between the NN predictions and the NLSE simulation results in both the time and frequency domains confirm the suitability of the NN to be used as a surrogate of the NLSE physical model.

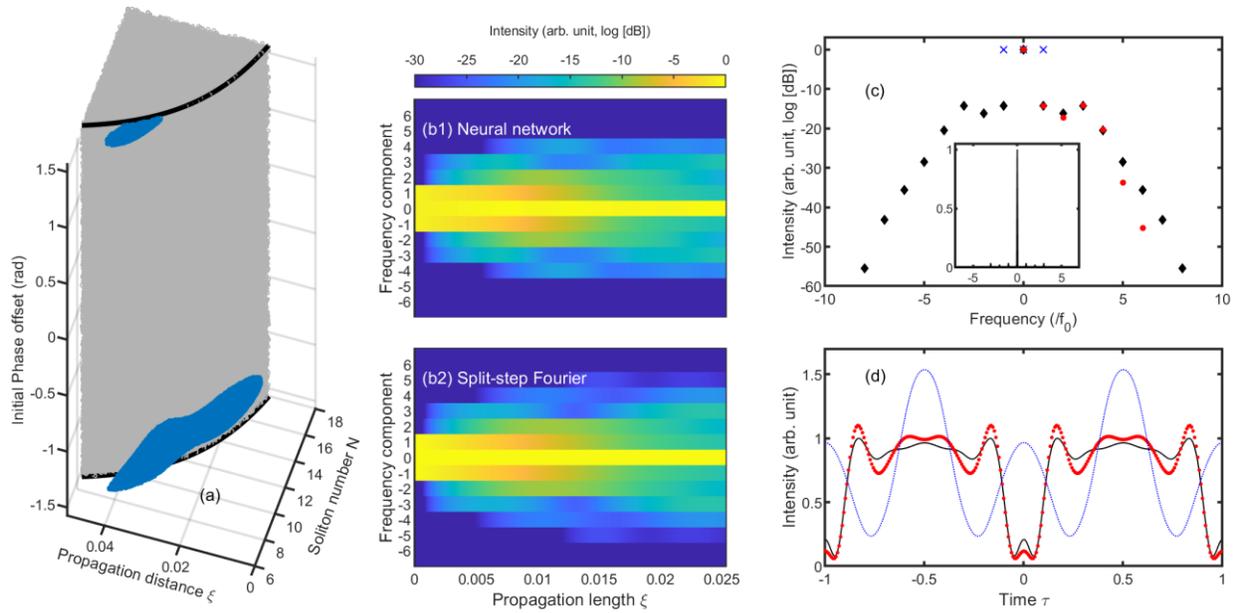

Figure 9: Spectral narrowing process. (a) Regions in the space of input parameters ($\varphi$, $\xi$, $N$) that enable focusing of more than 80% of the initial energy into the central frequency component, for an initial wave made of three initial spectral lines of equal amplitude and evolving in the normal dispersion regime of the fibre. (b) Longitudinal evolution of the optical spectrum obtained from the NN (panel 1) and NLSE simulations (panel 2). (c) Compressed optical spectrum plotted on a logarithmic scale. The predictions from the NN (red circles) are compared with the results of NLSE simulations (black diamonds). Also shown is the initial condition (blue crosses). The inset shows the compressed spectrum on a linear scale. (d) Corresponding temporal intensity profiles. The NN predictions, the results of NLSE simulations and the initial condition are plotted with red circles, a black curve, and a blue dotted curve, respectively. The waveforms are normalised to the same average power.

## C/ Undular bore patterns

As the last example, we investigate the spatiotemporal evolution of an initially sinusoidally modulated wave ($A = 3$dB and $\varphi = 0$) in both the anomalous and normal dispersion regimes of the fibre. In both cases, the initial $N$ number is set to 6 and a propagation length up to $\xi = 0.04$ is chosen. The results are summarised in Fig. 10, where we can observe the very different temporal dynamics experienced by the signal according to the regime of dispersion. In the anomalous dispersion regime (Fig. 10(a)), the signal pulses are significantly compressed, as evidenced by the temporal intensity profile recorded at the point of maximum compression $\xi = 0.029$. After this point, the pulses experience the typical dynamics observed for the breather solutions of the NLSE [42]. Conversely, in the presence of normal dispersion (Fig. 10(b)) the pulses incur very strong reshaping towards a parabolic form [8], followed by a stage of temporal broadening. Therefore, their wings start overlapping, leading to short-period oscillations. With further overlap, the number of oscillations increases and tends to undular bore patterns [43]. Once again, the results obtained from the NN agree well with those of direct NLSE simulations.

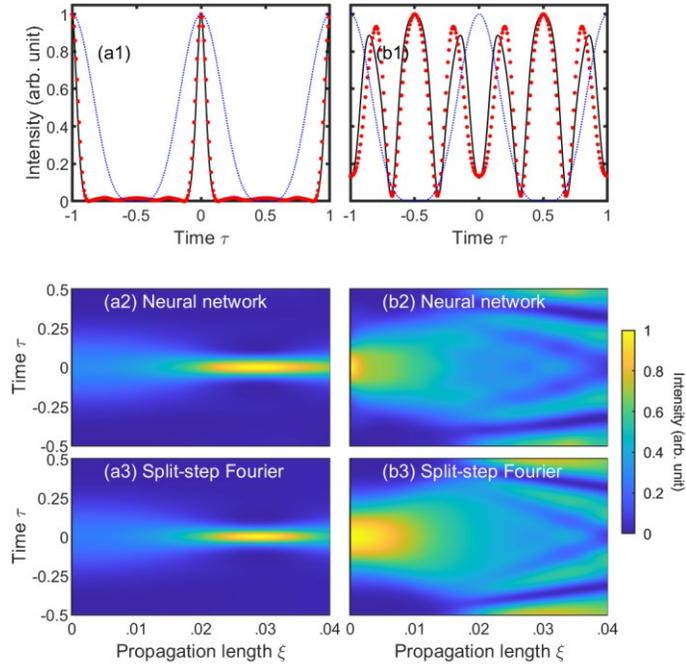

Figure 10: Temporal reshaping of an initially sinusoidal signal with $A = 3$ dB and $N = 6$ upon nonlinear propagation in the fibre. (a) Temporal narrowing in the anomalous dispersion regime and (b) formation of undular bores at normal dispersion. Panels 1 show the temporal intensity profiles at the point of maximum compression and after a propagation length of $\xi = 0.04$, respectively. The predictions from the NN (red circles) are compared with the results of NLSE simulations (black curves). Also shown is the initial signal (blue curves). The waveforms are normalised to the same peak power. Panels 2 and 3 show the longitudinal evolutions of the temporal intensity profiles as obtained from the NN and NLSE simulations, respectively.

# V. Conclusion

Following the great achievements of ML techniques in the field of photonics [44, 45], we have successfully trained an artificial NN to study the temporal and spectral evolutions of waves with three or four frequency components propagating in both the anomalous and normal dispersion regimes of a nonlinear optical fibre. Rather than considering the evolution of the temporal waveform as reported in many previous works, we have worked in the frequency domain, enabling significant reduction of the number of relevant output parameters. We have verified that the predictions from the trained NN show very close agreement with the results of direct numerical simulations of the NLSE for both the temporal and spectral intensity profiles. Although not explicitly shown in this work, as we have recorded both the modulus and phase of the spectral field, we were also fully able to reconstruct the phase and chirp profiles of the propagating waves. And we have also observed good agreement between the temporal chirp profiles predicted by the NN and those obtained from the NLSE model.

Taking advantage of the significant boost in computational speed enabled by the NN and using the underlying scaling rules of the propagation problem, we were able to probe the whole space of input parameters for solutions meeting pre-defined characteristics and to pinpoint the optimum parameter ranges. While the goal of the present paper is not to benchmark our search and optimisation strategy with evolutionary optimisation such as genetic algorithms [30], it is fully possible to combine the power of a genetic algorithm with the speed of a NN in an hybrid scheme. Benefitting from our NN-based surrogate model, we have demonstrated that starting from very simple initial wave conditions can lead to the formation of flat-top frequency combs in the fibre, thus eliminating the need for additional external line-by-line adjustments after propagation. We have also shown that compelling inverse four-wave mixing can occur for well-chosen initial parameters. We believe that the observation of these effective processes will stimulate new research to elucidate, based on physical arguments, the initial conditions that are required. We also anticipate applications of our results in the area of parametric optical amplification [46] and in the generation of high repetition-rate pulse trains with tailored intensity profiles. Indeed, whereas our initial studies have demonstrated that waves with three or four frequency components can be shaped into parabolic, triangular or rectangular-like pulse intensity profiles [18], the duty cycle of the resulting pulse trains could not be adjusted in those studies. Our new tool combined with a suitable merit function could help us identify the appropriate initial conditions for the generation of custom waveforms.

# Acknowledgements

We acknowledge the support of the Institut Universitaire de France (IUF). The project was funded by the Agence Nationale de la Recherche (OPTIMAL ANR-20-CE30-0004, EIPHI-BFC ANR-17-EURE-0002, I-SITE BFC ANR-15-IDEX-0003) and the CNRS (MITI interdisciplinary programs). The numerical simulations relied on the HPC resources of DNUM CCUB (Centre de Calcul de l'Université de Bourgogne). The authors also thank the GDR Elios (GDR 2080), and Bertrand Kibler for fruitful discussions.

# Declaration of Competing Interest

The authors declare that they have no known competing financial interests or personal relationships that could have appeared to influence the work reported in this paper.

# CRediT authorship contribution statement

**Christophe Finot:** Conceptualization, Formal analysis, Investigation, Software, Visualization, Writing-original draft, Funding acquisition. **Sonia Boscolo and John M. Dudley:** Validation and discussion, Writing - review & editing.

# Data availability

The data that support the findings of this study are available from the corresponding author, CF, upon reasonable request.